\documentclass[a4paper, 10pt]{article}

\usepackage[utf8]{inputenc}
\usepackage[T1]{fontenc}
\usepackage{geometry}
\usepackage[english]{babel}

\usepackage{amsmath, amssymb, amsthm}
\usepackage{hyperref}
\usepackage{graphicx}
\usepackage{subfigure}

\newtheorem{definition}{Definition}[section]
\newtheorem{theorem}[definition]{Theorem}
\newtheorem{proposition}[definition]{Proposition}
\newtheorem{lemma}[definition]{Lemma}

\newtheorem{example}[definition]{Example}

\title{{\Huge DRAFT}\\ Non-uniform cellular automata and distributions of rules}
\author{Julien Provillard \and Enrico Formenti \and Alberto Dennunzio}
\date{Laboratoire I3S\\ Universit\'e Nice Sophia Antipolis,\\ 2000, route des Lucioles - Les Algorithmes - b\^at. Euclide B - BP 12\\ 06903 Sophia Antipolis Cedex - France}

\begin{document}

\maketitle

\section{Introduction}

%TODO ajouter citations

Cellular automata (CA) are discrete dynamical systems consisting in an infinite number of finite automata arranged on a regular lattice. Each %O_O
node of the lattice contains a variable which can only take a finite number of different values. At any time, the state of the CA, also called its current configuration, is specified by the values of those variables. A CA make evolve its configuration in discrete time by applying a local rule simultaneously to all the variables in the lattice. The value of a variable in a new configuration is computed from the values of a finite number of variables of its neighborhood in the previous configuration according to the local rule. CA have been studied and used in a number of disciplines (computer science, mathematics, physic, biology, chemistry) with different purposes (representing natural phenomena, pseudo-random number generation, cryptography).

However the fact that a same rule is applied everywhere in the lattice can be a constraint in some cases (singular behavior, boundary conditions, non-resiliency). Then variants of CA have been introduced to allow the local rule used to compute the value of a variable to depend of its position. Those variants are called non-uniform cellular automata ($\nu$-CA) or hybrid cellular automata (HCA)  \cite{Dennunzio2011}.

In this paper we study $\nu$-CA on one-dimensional lattice defined over a finite set of local rules. The main goal is to determine how the local rules can be mixed to ensure the produced $\nu$-CA has some properties. In a first part, we give some background for the study of $\nu$-CA. Then surjectivity and injectivity are studied using a variant of DeBruijn graphs. The next part is dedicated to the number-conserving property. In a last section, we will be interested in the dynamical properties for linear $\nu$-CA.

\section{Definitions}

\subsection{Automata}

\begin{definition}[Configuration]
Let $A$ be a finite set ($|A| \geq 2$). A (one-dimensional) configuration over $A$ is a mapping from $\mathbb{Z}$ to $A$. If $c$ is a configuration and $i$ an integer $c(i)$ is the state of the configuration at index $i$, this state is more often written $c_i$ to simplify notation.
\end{definition}

$A$ is called the alphabet of the configuration. $\mathcal{C}_A$ denotes the set of all configurations on the alphabet $A$. For a configuration $c$, $c_{[i,j]}$ denotes the sequence of states $c_i c_{i+1} \dots c_j$.

A local rule of radius $r$ on the alphabet $A$ is a mapping from $A^{2r+1}$ to $A$. $\mathcal{R}_{A,r}$ denotes the set of all local rules of radius $r$ on $A$ and $\mathcal{R}_A = \cup_{r \in \mathbb{N}} \mathcal{R}_{A,r}$ the set of all local rules on $A$. Local rules are central in the definition both of cellular automata and non-uniform cellular automata.

\begin{definition}[CA]
\label{def_CA}
A mapping $F : \mathcal{C}_A \rightarrow \mathcal{C}_A$ is a cellular automaton if
$$\exists r \in \mathbb{N}, \exists f \in \mathcal{R}_{A,r}, \forall i \in \mathbb{Z}, \forall x \in \mathcal{C}_A, H(x)_i = f(x_{[i-r,i+r]}) \enspace.$$
\end{definition}

\begin{definition}[$\nu$-CA]
\label{def_NuCA}
A mapping $F : \mathcal{C}_A \rightarrow \mathcal{C}_A$ is a non-uniform cellular automaton if
$$\forall i \in \mathbb{Z}, \exists r \in \mathbb{N}, \exists f \in \mathcal{R}_{A,r}, \forall x \in \mathcal{C}_A, H(x)_i = f(x_{[i-r,i+r]}) \enspace.$$
\end{definition}

\begin{definition}[r$\nu$-CA]
\label{def_NuCA}
A mapping $F : \mathcal{C}_A \rightarrow \mathcal{C}_A$ is a non-uniform cellular automaton with fixed radius if
$$\exists r \in \mathbb{N}, \forall i \in \mathbb{Z}, \exists f \in \mathcal{R}_{A,r}, \forall x \in \mathcal{C}_A, H(x)_i = f(x_{[i-r,i+r]})\enspace.$$
\end{definition}

The definition \ref{def_CA} is the classical definition for CA. A CA is fully determined by its local rule which is applied simultaneously at all sites. The definition of $\nu$-CA allows different rules at different sites, each rule can have its own radius but all share the same alphabet. This generic definition is quite strong. Indeed even if each rule accesses to a finite number of data of the configuration, this number can be unbounded. This behavior is not expected in classical studies because the notion of locality makes less sense. The r$\nu$-CA are an intermediary model which allow different rules at different sites but each rule accesses to the same range of data according to its position.

\begin{example}[Shift automaton]
\label{shift}
The shift automaton $\sigma$ on an alphabet $A$ is defined as follow
$$\forall i \in \mathbb{Z}, \forall x \in \mathcal{C}_A, \sigma(x)_i = x_{i+1} \enspace.$$
It is a CA of local rule $f$ where
$$
\begin{array}{rccl}
f : & A^3 & \rightarrow & A \\
 & (x,y,z) & \rightarrow & z\\
\end{array}\enspace.
$$
\end{example}

\subsection{Distributions}

\begin{definition}[Distribution]
Let $\mathcal{R}$ be a subset of $\mathcal{R}_A$, a distribution on $\mathcal{R}$ is an application $\theta$ from $\mathbb{Z}$ to $\mathcal{R}$. As for configurations, if $\theta$ is a distribution and $i$ an integer $\theta(i)$, or $\theta_i$, is the local rule of the distribution at index $i$.
\end{definition}

$\Theta_{\mathcal{R}}$ denotes the set of all distributions on $\mathcal{R}$. For a distribution $\theta$, $\theta_{[i,j]}$ denotes the sequence of local rules $\theta_i \theta_{i+1} \dots \theta_j$.

A distribution of rules $\theta$ induces a $\nu$-CA $H_{\theta}$ defined by

$$\forall i \in \mathbb{Z}, \forall x \in \mathcal{C}_A, H_{\theta}(x)_i = \theta_i(x_{[i-r_i,i+r_i]})$$

where $r_i$ is the radius of the rule $\theta_i$.

\begin{proposition}
\label{finite_number_rules}
If $\mathcal{R}$ is finite, then for all distribution $\theta$ on $\mathcal{R}$, $H_{\theta}$ is a r$\nu$-CA.
\end{proposition}

\begin{proof}
Let denote $r = \max \{ n \in \mathbb{N} : f \in \mathcal{R} \cap \mathcal{R}_{A,n}\}$ the greatest radius of a rule in $\mathcal{R}$. Then for a rule $f \in \mathcal{R}$ of radius $r_f$, we define a rule $\tilde{f}$ of radius $r$ by
$$ \tilde{f}(x) = f(x_{[r-r_f, r + r_f]}) \enspace.$$

Let $\theta$ be a distribution of $\mathcal{R}$ and $H_{\theta}$ the $\nu$-CA induced by $\theta$,
$$\forall i \in \mathbb{Z}, \forall x \in \mathcal{C}_A, H_{\theta}(x)_i = \theta_i(x_{[i-r_i,i+r_i]} = \tilde{\theta_i}(x_{[i-r,i+r]})$$

where $r_i$ is the radius of the rule $\theta_i$. Then $H_{\theta}$ is r$\nu$-CA of radius $r$.

\end{proof}

In this paper, we will consider distributions on a finite set $\mathcal{R}$ of local rules. The proof of the Proposition \ref{finite_number_rules} shows we can always assume that $\mathcal{R}$ is a subset of $\mathcal{R}_{A,r}$ for an integer $r$. Moreover, each finite distribution $\psi$ (finite sequence of rules in $\mathcal{R}$) of $n$ rules defines a function $h_{\psi} : A^{n+2r} \rightarrow A^n$ by
$$\forall x \in A^{n+2r}, \forall i \in \{0, \dots, n-1\}, h_{\psi}(x)_i = \psi_i(x_{[i,i+2r]}) \enspace.$$
These functions are called partial transition functions since they express the behavior of a $\nu$-CA on a finite set of sites : if $\theta$ is a distribution and $i \leq j$ are integers, then
$$\forall x \in \mathcal{C}_A, H_{\theta}(x)_{[i,j]} = h_{\theta_{[i,j]}(x_{[i-r,j+r]})} \enspace.$$

\subsection{Topological dynamics}

The topological properties for $\nu$-CA are studied according to the Cantor distance $d$. For two configurations $x$ and $y$, the Cantor distance is defined by
$$d(x,y) =
\left\{
\begin{array}{cl}
0 & \text{if $x = y$} \\
2^{-k} & \text{ where $k = \min \{ i \in \mathbb{N} : x_{[-i,i]} \neq y_{[-i,i]}\}$, otherwise} \\
\end{array}
\right.
\enspace.
$$

The topology defined by the Cantor distance on $\mathcal{C}_A = A^{\mathbb{Z}}$ coincide with the product topology induced by the discrete topology on $A$. Then the space of configurations $(\mathcal{C}_A, d)$ is a Cantor space : it is a perfect, compact, totally disconnected metric space.

Cellular automata and non-uniform cellular automata can be characterized by this topology.

\begin{proposition}
Let $H : \mathcal{C}_A \rightarrow \mathcal{C}_A$ be a function, then
\begin{enumerate}
\item
$H$ is a CA if and only if $H$ is continuous and commutes with the shift automaton, i.e. $H \circ \sigma = \sigma \circ H$ \cite{Hedlund1969}.
\item
$H$ is a $\nu$-CA if and only if $H$ is continuous.
 \item
If $H$ is a r$\nu-CA$ then $H$ is Lipschitz continuous.
\end{enumerate}
\end{proposition}

However, the following example shows that there exists $\nu$-CA which are Lipschitz continuous but are not r$\nu$-CA.

\begin{example}
The $\nu$-CA $H$ defined on an alphabet $A$ as
$$\forall i \in \mathbb{Z}, \forall x \in \mathcal{C}_A, H(x)_i = x_{-i}$$
is Lipschitz continuous (with a Lipschitz constant equal to 1) but it is not a r$\nu$-CA.
\end{example}

In Section \ref{eq_sens_add}, we will study some topological properties of a subclass of $\nu$-CA, namely equicontinuity and sensitivity to initial conditions.

\begin{definition}[Equicontinuity point]
A configuration $x \in \mathcal{C}_A$ is said to be an equicontinuity point of the function $H : \mathcal{C}_A \rightarrow \mathcal{C}_A$ if and only if
$$\forall \epsilon > 0, \exists \delta > 0, \forall y \in \mathcal{C}_A, d(x,y) < \delta \Rightarrow \forall n \in \mathbb{N}, d(H^n(x),H^n(y)) < \epsilon \enspace.$$
\end{definition}

\begin{definition}[Equicontinuity]
A function $H : \mathcal{C}_A \rightarrow \mathcal{C}_A$ is said to be equicontinuous if and only if all the configurations are equicontinuous points.
\end{definition}

\begin{definition}[Sensitivity to initial conditions]
A function $H : \mathcal{C}_A \rightarrow \mathcal{C}_A$ is said to be sensitive to initial conditions (or just sensitive) if and only if
$$\exists \epsilon > 0, \forall x \in \mathcal{C}_A, \forall \delta > 0, \exists y \in \mathcal{C}_A, d(x,y) < \delta \text{ and } \exists n \in \mathbb{N}, d(H^n(x), H^n(y)) > \epsilon \enspace.$$
\end{definition}

Equicontinuity is a property of stability of the system, while sensitivity to initial conditions is more related to chaotic behavior.

\section{Surjectivity and injectivity}

Let $\mathcal{R}$ be a finite set of rules. We assume, without loss of generality, that all rules of $\mathcal{R}$ have same radius $r$. We want to determine which are the distributions of $\Theta_{\mathcal{R}}$ inducing surjective (resp. injective) r$\nu$-CA.

\subsection{Surjectivity}

Let $Surj(\mathcal{R}) = \{ \theta \in \Theta_{\mathcal{R}} : H_{\theta} \text{ is surjective}\}$ denote the set of all distributions inducing surjective r$\nu$-CA. We are going to prove that $Surj(\mathcal{R})$ is a sofic subshift.

Recall that a subset $U$ of distributions is a subshift if it is (topologically) closed and shift invariant, i.e. $\sigma(U) = U$. Equivalently, a subshift $U$ can be defined by a set of forbidden patterns. Then a distribution $\theta$ is in $U$ if and only if no finite pattern of $\theta$ is forbidden. A subshift is said to be sofic if it can be defined by a set of forbidden patterns which is recognizable by a finite automaton \cite{Lind1995}.

The proof consists in three steps. First we show that a r$\nu$-CA induced by a distribution is surjective if and only if all its partial transition functions are surjective. Second we prove that  $Surj(\mathcal{R})$ is the subshift that avoids the set of all non-surjective partial transition functions on $\mathcal{R}$. Finally, we show that this set is recognizable by a finite automaton, and hence the subshift is sofic.

\begin{proposition}
Let $\theta \in \Theta_{\mathcal{R}}$, $H_{\theta}$ is surjective if and only if for all $i \leq j$, $h_{\theta_{[i,j]}}$ is surjective.
\end{proposition}

\begin{proof}
Assume $H_{\theta}$ is surjective, let $i \leq j$ be two integers and $w \in A^{j-i+1}$. Choose a configuration $x$ such that $x_{[i,j]} = w$. Since $H_{\theta}$ is surjective, there exists $y$ such that $H_{\theta}(y) = x$. $h_{\theta_{[i,j]}}(y_{[i-r,j+r]}) = H_{\theta}(y)_{[i,j]} = x_{[i,j]} = w$. For all $i \leq j$, $h_{\theta_{[i,j]}}$ is surjective.

\medskip
Assume for all $i \leq j$, $h_{\theta_{[i,j]}}$ is surjective. Let $x$ be a configuration and, for all integer $n \geq 0$, let $Y_n$ be the set $\{ y \in \mathcal{C}_A : H_{\theta}(y)_{[-n,n]} = x_{[-n,n]}\}$.

For all $n \geq 0$, $h_{\theta_{[-n,n]}}$ is surjective, then there exists $w \in A^{2(n+r) + 1}$ such that $h_{\theta_{[-n,n]}}(w) = x_{[-n,n]}$. Then all configuration $y$ such that $y_{[-n-r,n+r]} = w$ are in $Y_n$, and $Y_n$ is not empty.

Let $n \geq 0$ be an integer and $y \in Y_{n+1}$, then $H_{\theta}(y)_{[-n-1,n+1]} = x_{[-n-1,n+1]}$ and $y \in Y_n$.

For all $n$, $Y_n \neq \emptyset$ and $Y_n \subseteq Y_{n+1}$, by compacity there exists $y \in \cap_{n \geq 0} Y_n$. Such an $y$ verifies $H_{\theta}(y) = x$ and therefore $H_{\theta}$ is surjective.
\end{proof}

Let $\mathcal{F}_{\mathcal{R}} = \{ \psi \in \mathcal{R}^* : h_{\psi} \text{ is not surjective} \}$ denotes the set of all finite distributions which define non-surjective partial transition functions. Then a distribution $\theta \in \Theta_{\mathcal{R}}$ defines a surjective r$\nu$-CA if and only if $\theta$ avoids the patterns of $\mathcal{F}_{\mathcal{R}}$. $Surj(\mathcal{R})$ is the subshift of $\theta_{\mathcal{R}}$ that avoids $\mathcal{F}_{\mathcal{R}}$.

Recall that the decidability of the surjectivity for classical cellular automata has been studied thanks to DeBruijn graphs \cite{Durand1998}. We will present here a variant of those graphs for distributions on a finite set of rules. The study of this graph will allow us to show that $\mathcal{F}_{\mathcal{R}}$ is recognizable.

The DeBruijn graph associated to the set of rules $\mathcal{R}$ (we assume the radius of the rules is at least 1) is the graph $\mathcal{G}_{\mathcal{R}}$ which contains $|A|^{2r}$ nodes, each of them is labeled by a different word of $A^{2r}$. For every states $a$ and $b$ of $A$, for every word $w$ of $A^{2r-1}$, for every rule $f$ of $\mathcal{R}$, there exists an edge from the node $aw$ to the node $wb$ (nodes are assimilated to their label). This edge is labeled by $(f,f(awb))$.

\begin{example}
Let $A = \{0,1\}$ and $\mathcal{R} = \{\oplus, id\}$ where
$$
\begin{array}{rccl}
\oplus : & A^3 & \rightarrow & A \\
 & (x,y,z) & \rightarrow & x + z \bmod 2\\
\\
id : & A^3 & \rightarrow & A \\
 & (x,y,z) & \rightarrow & y\\
\end{array}
\enspace.
$$
The DeBruijn graph $\mathcal{G}_{\mathcal{R}}$ associated to $\mathcal{R}$ is the graph
\begin{center}
\includegraphics{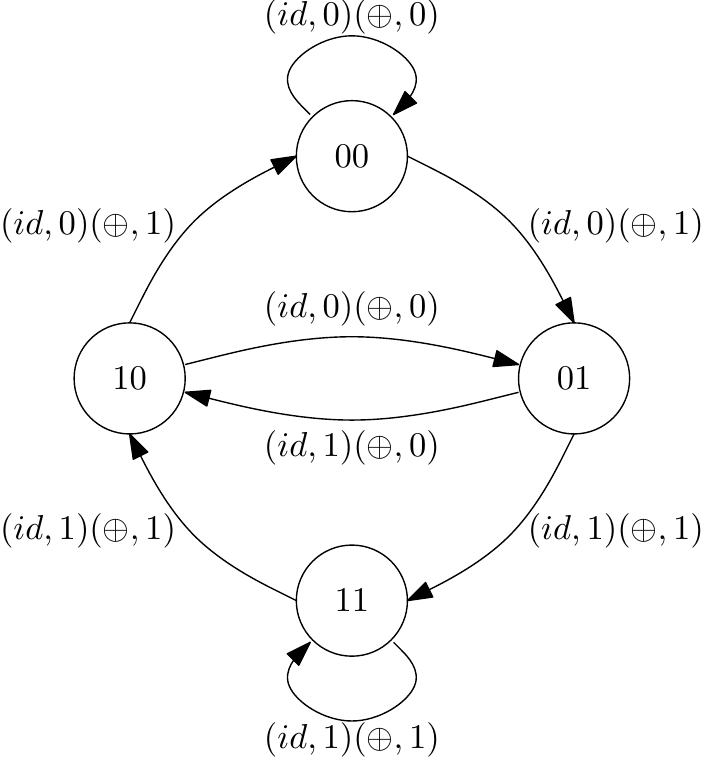}
\end{center}
where multiple edges from same origin to same target have been collapsed and labels concatenated.
\end{example}

Consider this graph as a finite automaton where all states are both initial and final and let $\mathcal{L}$ be the language recognized by this automaton.

\begin{lemma}
\label{language}
$\mathcal{L} = \{(\psi,u) \in (\mathcal{R} \times A)^* : h_\psi^{-1}(u) \neq \emptyset \}$.
\end{lemma}

\begin{proof}
Let $(\psi,u) \in (\mathcal{R} \times A)^*$ be such that $h_\psi^{-1}(u) \neq \emptyset$ and $n = |\psi| = |u|$. There exists $w \in A^{n+2r}$ such that $h_\psi(w) = u$. Indeed,
$$w_{[0,2r-1]} \xrightarrow{(\psi_0,u_0)} w_{[1,2r]} \xrightarrow{(\psi_1,u_1)} w_{[2,2r+1]} \xrightarrow{(\psi_2,u_2)} \dots \xrightarrow{(\psi_{n-1},u_{n-1})} w_{[n,n+2r-1]}$$
is a path of $\mathcal{G}_{\mathcal{R}}$ and then $(\psi,u) \in \mathcal{L}$.

\medskip
Let $(\psi,u)$ be in $\mathcal{L}$, $(\psi,u)$ is the label of a path in $\mathcal{G}_{\mathcal{R}}$ of the form
$$a_0 w_0 \xrightarrow{(\psi_0,u_0)} w_0b_1 = a_1 w_1 \xrightarrow{(\psi_1,u_1)} w_1b_2 = a_2 w_2 \xrightarrow{(\psi_2,u_2)} \dots \xrightarrow{(\psi_{n-1},u_{n-1})} w_{n-1}b_n = a_n w_n$$
where for all $i$, $a_i$ and $b_i$ in $A$ and $w_i$ in $A^{2r-1}$. Then by definition of $\mathcal{G}_{\mathcal{R}}$, $h_{\psi}(a_0\dots a_n w_n) = u$ and $h_{\psi}^{-1}(u) \neq \emptyset$.
\end{proof}

\begin{proposition}
$Surj(\mathcal{R})$ is a sofic subshift.
\end{proposition}

\begin{proof}
We have a finite automaton that recognizes $\mathcal{L} = \{(\psi,u) \in (\mathcal{R} \times A)^* : h_\psi^{-1}(u) \neq \emptyset \}$ (Lemma \ref{language}). Then we can construct an automaton $\mathcal{A}$ which recognizes $\mathcal{L}^c = \{(\psi,u) \in (\mathcal{R} \times A)^* : h_\psi^{-1}(u) = \emptyset \}$. We construct a new automaton $\tilde{\mathcal{A}}$ from $\mathcal{A}$ by deleting all the second components of edge labels. A word $\psi \in \mathcal{R}^*$ is recognized by $\tilde{\mathcal{A}}$ if and only if there exists $u \in A^*$ such that $(\psi,u) \in \mathcal{L}^c$, i.e. $h_\psi^{-1}(u) = \emptyset$. Finally, a word $\psi \in \mathcal{R}^*$ is recognized by $\tilde{\mathcal{A}}$ if and only if $h_{\psi}$ is not surjective and the language recognized by $\tilde{\mathcal{A}}$ is $\mathcal{F}_{\mathcal{R}}$. $Surj(\mathcal{R})$ is a sofic subshift.
\end{proof}

\subsection{Injectivity}

We have seen that paths in the DeBruijn graph $\mathcal{G}_{\mathcal{R}}$ associated to a set of rules $\mathcal{R}$ allow to define a word $w$ (from the sequence of visited nodes) and a finite distribution $\psi$ and another word $u$ (from the sequence of visited edges) such that $h_{\psi}(w) = u$. If we consider now bi-infinite paths, we defined in the same way two configurations $x$ and $y$ and a distribution $\theta$ such that $H_{\theta}(x) = y$. The product graph of $\mathcal{R}$ is an extension of its DeBruijn graph which allows to describe simultaneously two configurations which have the same image.

The product graph $\mathcal{P}_{\mathcal{R}}$ is the graph which contains $|A|^{4r}$ nodes, each of them is labeled by a different couple of word of $A^{2r}$. There is an edge from $(u,u')$ to $(v,v')$ labeled by $(f,a) \in \mathcal{R} \times A$ if and only if there is an edge from $u$ to $v$ (resp. from $u'$ to $v'$) labeled by $(f,a)$.

Then a bi-infinite path in $\mathcal{P}_{\mathcal{R}}$ corresponds to two bi-infinite paths in $\mathcal{G}_{\mathcal{R}}$ and the sequence of edge labels used by these two paths is the same. More formally, if
$$\dots \xrightarrow{(\theta_{i-1},z_{i-1})} (u_{i},u'_{i}) \xrightarrow{(\theta_i,z_i)} (u_{i+1},u'_{i+1}) \xrightarrow{(\theta_{i+1},z_{i+1})} (u_{i+2},u'_{i+2}) \xrightarrow{(\theta_{i+2},z_{i+2})} \dots$$
is a path in $\mathcal{P}_{\mathcal{R}}$, define two configurations $x$ and $y$ such as for all integer $i$, $x_i$ is the $(r + 1)^{th}$ letter of $u_i$ and $y_i$ is the $(r + 1)^{th}$ letter of $u'_i$. Then $H_{\theta}(x) = H_{\theta}(y) = z$.

On the other hand, if $\theta$ is a distribution and $x$ and $y$ are two configurations such that $H_{\theta}(x) = H_{\theta}(y)$ = z, then 
$$\dots \xrightarrow{(\theta_{i-1},z_{i-1})} (x_{[i-r,i+r-1]},y_{[i-r,i+r-1]}) \xrightarrow{(\theta_i,z_i)} (x_{[i-r+1,i+r]},y_{[i-r+1,i+r]}) \xrightarrow{(\theta_{i+1},z_{i+1})} \dots$$
is a path in $\mathcal{P}_{\mathcal{R}}$.

As for the surjectivity, denote $\tilde{\mathcal{P}}_{\mathcal{R}}$ the finite automaton obtained from $\mathcal{P}_{\mathcal{R}}$ by deleting all the second components of edge labels. A bi-infinite path defines two configurations $x$ and $y$ (from the sequence of visited nodes) and a configuration $\theta$ (from the sequence of visited edges) such that $H_{\theta}(x) = H_{\theta}(y)$. The converse is also trivially true.

Then $\theta$ leads to a non-injective r$\nu$-CA $H_ {\theta}$ if and only there exists two configurations $x \neq y$ such that $H_{\theta}(x) = H_{\theta}(y)$ if and only if there exists a bi-infinite path in $\tilde{\mathcal{P}}_{\mathcal{R}}$ labeled by $\theta$ which visits a node $(u,u')$ where $u \neq u'$.

Recall that a $\zeta$-rational language is a set of bi-infinite words recognized by a finite automaton, that is to say the set of all labels of successful bi-infinite paths in the automaton. A path in such automaton is successful if and only it crosses infinitely many times initial states for negative indexes and infinitely many times final states for positive indexes (Büchi acceptance condition).

\begin{proposition}
$Inj(\mathcal{R}) = \{ \theta \in \Theta_{\mathcal{R}} : H_{\theta} \text{ is injective}\}$, i.e. the set of all distributions that induce injective r$\nu$-CA, is a $\zeta$-rational language.
\end{proposition}

\begin{proof}
Consider now $\tilde{\mathcal{P}}_{\mathcal{R}}$ as a finite automaton where all the states are initial and the final states are the states of the form $(u,u')$ with $u \neq u'$. A bi-infinite path is successful in this graph if and only it crosses infinitely many times initial states for negative indexes (automatically because all states are initial) and at least one final state. Then the language recognized (under this acceptance condition) is exactly $\Theta_{\mathcal{R}} \smallsetminus Inj(\mathcal{R})$.

But if a language is recognized by an automaton with this new acceptance condition, it is recognized by an automaton with the Büchi acceptance condition and then is $\zeta$-rational. It is known that $\zeta$-rational are closed under complement, since $\Theta_{\mathcal{R}} \smallsetminus Inj(\mathcal{R})$ is 
$\zeta$-rational then $Inj(\mathcal{R})$ is $\zeta$-rational.
\end{proof}

\section{Number conserving}

In physics, a lot of transformations are conservative : a certain quantity remains invariant during a whole experiment. Think to conservation laws of mass and energy for example. As $CA$ and $\nu$-CA are used to represent phenomena from physics, such a property of conservation has been introduced. We will generalize some existing results of the uniform case.

\medskip
In this section, the alphabet $A$ we consider is a "numerical" alphabet of the form $\{0, 1, \dots, s-1\}$ where $s$ is the cardinal of $A$. A configuration $x \in \mathcal{C}_A$ is said to be \emph{finite} if and only if the support of $x$, i.e. the set $\{i \in \mathbb{Z}, x_i \neq 0\}$, is finite. Let $\mathcal{C}_A^F$ denotes the set of all finite configurations on $A$ and $\underline{0}$ be the configuration with empty support, i.e. such that for all integer $i$, $\underline{0}_i = 0$.

For every configuration $x \in \mathcal{C}_A$, define the \emph{partial charge} of $x$ between the index $-n$ and $n$ by
$$\mu_n(x) = \sum_{i = -n}^n x_i$$
and the \emph{global charge} of $x$ by
$$\mu(x) = \lim_{n \rightarrow \infty} \mu_n \enspace.$$
Then, if $x$ is not a finite configuration, $\mu(x) = \infty$.

\begin{definition}[FNC]
A $\nu$-CA $H$ is \emph{number-conserving on finite configuration} (FNC) if and only if for all $x \in \mathcal{C}_A^F$, $\mu(x) = \mu(H(x))$.
\end{definition}

A $\nu$-CA $H$ is FNC if and only if it preserves the charge of finite configurations. Consequently, for all finite configuration $x$, $H(x)$ is a finite configuration and $H(\underline{0}) = \underline{0}$.

In the case of non-finite configurations, the conservation of the charge is expressed as conservation of the average charge. The average charge of a configuration $x$ on a window of size $n$ is the quantity $\frac{\mu_n(x)}{2n+1}$. Then a $\nu$-CA $H$ will be number-conserving if the average charge of a configuration and the average charge of its image are asymptotically the same
$$\forall x \in \mathcal{C}_A, \frac{\mu_n(H(x))}{2n+1} \sim \frac{\mu_n(x)}{2n+1} \enspace.$$

Then if $\mu_n(x) \neq 0$ the quantity $\frac{\mu_n(H(x))}{\mu_n(x)}$ represents the relative gain/loss of charge on a window of size $2n+1$.

\begin{definition}[NC]
Let $H$ be a $\nu$-CA and $x$ be a configuration. If $x \neq \underline{0}$ then there exists $n_0$ such that for all $n$ greater than or equals to $n_0$, $\mu_n(x) \neq 0$. Then the sequence $\frac{\mu_n(H(x))}{\mu_n(x)}$ is defined for $n$ greater than or equals to $n_0$. Let $m(x) = \liminf_{n \rightarrow \infty} \frac{\mu_n(H(x))}{\mu_n(x)}$ and $M(x) = \limsup_{n \rightarrow \infty} \frac{\mu_n(H(x))}{\mu_n(x)}$.

$H$ is said to be number-conserving (NC) if and only if
\begin{enumerate}
\item
$H(\underline{0}) = \underline{0}$
\item
$\forall x \in \mathcal{C}_A \smallsetminus \{\underline{0}\}, m(x) = M(x) = 1$ (the sequence $\frac{\mu_n(H(x))}{\mu_n(x)}$ converges on 1).
\end{enumerate}
\end{definition}

This definition of number-conserving $\nu$-CA does not depend from the origin chosen for the lattice. In fact, the sequences $\frac{\mu_n(H(x))}{\mu_n(x)}$ and $\frac{\mu_n(H(\sigma(x)))}{\mu_n(\sigma(x))}$ have the same adherence values for all configuration $x$. Then
$$\forall x \in \mathcal{C}_A, m(x) = m(\sigma(x)) \text{ and } M(x) = M(\sigma(x)) \enspace.$$

\begin{proposition}
Let $H$ be a r$\nu$-CA of radius $r$, $H$ is NC if and only if $H$ is $NFC$.
\end{proposition}

\begin{proof}
Assume that $H$ is NC. Since $H(\underline{0}) = \underline{0}$, all images of finite configurations are finite configurations. Let $x$ be in $\mathcal{C}_A^F \smallsetminus \{\underline{0}\}$, the sequences $\mu_n(x)$ and $\mu_n(H(x))$ are stationary and converge respectively on $\mu(x)$ and $\mu(H(x))$. Then $\lim_{n \rightarrow \infty} \frac{\mu_n(H(x))}{\mu_n(x)} = \frac{\mu(H(x))}{\mu(x)} = 1$. $\mu(x) = \mu(H(x))$ and $H$ is FNC.

\medskip
Assume that $H$ is not NC. If $H(\underline{0}) \neq \underline{0}$ then $H$ is not FNC, else there exists a configuration $x$ such that $m(x) \neq 1$ or $M(x) \neq 1$. If $x$ is a finite configuration then $H$ is not FNC.

We assume now that $x$ is not a finite configuration ($\mu(x) = \infty$). We will now be interested in the case $M(x) > 1$ (the other case $m(x) < 1$ has a similar proof). By definition of upper limit, for all $\epsilon > 0$, there exists an infinite number of indexes $n$ such that

$$\frac{\mu_n(H(x))}{\mu_n(x)} \geq M(X) - \epsilon \enspace.$$

Let $\epsilon > 0$ and $k > 0$, there exists infinitely many indexes $n$ such that
\begin{equation}
\frac{\mu_n(H(x))}{\mu_n(x)} \geq M(X) - \frac{\epsilon}{2}
\end{equation}
which is equivalent to say that there exists infinitely many indexes $n$ such that
\begin{equation}
\label{eq_k}
\mu_n(H(x)) \geq (M(X) - \epsilon) \mu_n(x) + \frac{\epsilon}{2} \mu_n(x)\enspace.
\end{equation}

$\lim_{n \rightarrow \infty} \mu_n(x) = \infty$ then there exists $n_0$ such that for all integer $n > n_0$, $\mu_n(x) \geq \frac{2k}{\epsilon}$. But as there are infinitely many indexes that verify inequality (\ref{eq_k}), one of them is greater than $n_0$ and for this index $n$ we have
\begin{equation}
\mu_n(H(x)) \geq (M(X) - \epsilon) \mu_n(x) + k \enspace.
\end{equation}

We have proved that
$$\forall \epsilon > 0, \forall k > 0, \exists n, \mu_n(H(x)) \geq (M(X) - \epsilon) \mu_n(x) + k \enspace.$$

Choose $\epsilon$ such that $M(x) - \epsilon > 1$ and $k = 2r(s-1)$, then there exists an integer $n$ such that
$$\mu_n(H(x)) \geq (M(X) - \epsilon) \mu_n(x) + 2r(s-1) \enspace.$$

Let $y$ be the finite configuration such that for all integer $i$, $y_i = x_i$ if $|i| \leq n$, 0 otherwise.

$$
\begin{array}{rcl}
\mu(H(y)) & = & \sum_{i \in \mathbb{Z}} H(y)_i\\
 & = & \sum_{i = -n-r}^{n+r} H(y)_i \\
 & \geq & \sum_{i = -n+r}^{n-r} H(x)_i \\
 & \geq & \sum_{i = -n}^n H(x)_i - 2r(s-1) \\
 & \geq & (M(x)-\epsilon)\sum_{i = -n}^n x_i \\
 & \geq & (M(x)-\epsilon)\sum_{i \in \mathbb{Z}} y_i\\
 & > & \mu(y) \enspace.
\end{array}
$$
Hence $H$ is not FNC.
\end{proof}

\paragraph{Application to r$\nu$-CA defined on a finite set of rules.}

Let $\mathcal{R}$ be as usual a finite set of local rules of radius $r> 0$. Let $NC(\mathcal{R}) = \{ \theta \in \Theta_{\mathcal{R}} : H_{\theta} \text{ is NC}\}$ be the set of all distributions that induce number-conserving r$\nu$-CA. We will prove that $NC(\mathcal{R})$ is a subshift of finite type.

\begin{lemma}
Let $\theta \in \Theta_{\mathcal{R}}$. Then $\theta \in NC(\mathcal{R})$ if and only if $\forall j \in \mathbb{Z}, \theta_{[j-2r,j]} \notin \mathcal{F}_{\mathcal{R}}$ where
$$\mathcal{F}_{\mathcal{R}} = \{\psi \in \mathcal{R}^{2r+1} : \exists u \in A^{2r+1}, \psi_{2r}(u) \neq u_0 + \sum_{i=0}^{2r-1} \psi_{i+1}(0^{2r-i}u_{[1,i+1]}) - \psi_{i}(0^{2r-i}u_{[0,i]})\} \enspace.$$
\end{lemma}

\begin{proof}
Assume that $\theta$ is in $NC(\mathcal{R})$, let $j \in \mathbb{Z}$ and $u \in A^{2r+1}$. $H_{\theta}$ is NC then $H_{\theta}(\underline{0}) = \underline{0}$ and for all integer $i$, $\theta_i(0^{2r+1}) = 0$.

Let $x$ be the finite configuration such that $x_{[j-r,j+r]} = u$ and $x_i = 0$ elsewhere, let $y$ be the finite configuration such that $y_{[j-r,j+r]} = 0u_{[1,2r]}$ and $y_i = 0$ elsewhere.

$H_{\theta}$ is NC (and FNC) then $\mu(H(x)) = \mu(x)$, i.e.

\begin{equation}
\label{eq_u}
\sum_{i=0}^{2r} \theta_{j+i-2r}(0^{2r-i}u_{[0,i]}) + \sum_{i=1}^{2r} \theta_{j+i}(u_{[i,2r]}0^i) = \sum_{i = 0}^{2r} u_i \enspace.
\end{equation}

As same $\mu(H(y)) = \mu(y)$, i.e.

\begin{equation}
\label{eq_0u}
\sum_{i=1}^{2r} \theta_{j+i-2r}(0^{2r-i+1}u_{[1,i]}) + \sum_{i=1}^{2r} \theta_{j+i}(u_{[i,2r]}0^i) = \sum_{i = 1}^{2r} u_i \enspace.
\end{equation}

Subtracting (\ref{eq_0u}) to (\ref{eq_u}), we obtain

\begin{equation}
\theta_j(u) = u_0 + \sum_{i=1}^{2r} \theta_{j+i-2r}(0^{2r-i+1}u_{[1,i]}) - \sum_{i=0}^{2r-1} \theta_{j+i-2r}(0^{2r-i}u_{[0,i]})
\end{equation}

which can be rewritten

\begin{equation}
\theta_j(u) = u_0 + \sum_{i=0}^{2r-1} \theta_{j+i+1-2r}(0^{2r-i}u_{[1,i+1]}) - \theta_{j+i-2r}(0^{2r-i}u_{[0,i]}) \enspace.
\end{equation}

That is true for all word $u$ then $\theta_{[j-2r,j]} \notin \mathcal{F}_{\mathcal{R}}$.

\medskip
Assume that for all integer $j$, $\theta_{[j-2r,j]} \notin \mathcal{F}_{\mathcal{R}}$. Let $j$ be an integer, $\theta_{[j,j+2r]} \notin \mathcal{F}_{\mathcal{R}}$ then, taking $u = 0^{2r+1}$, we have
$$\theta_{j + 2r}(0^{2r+1}) = 0 + \sum_{i=0}^{2r-1} \theta_{j + i +1}(0^{2r+1}) - \theta_{j + i}(0^{2r+1})$$
which leads to $\theta_j(0^{2r+1}) = 0$. This will justify that all following sums have in fact a finite support and are well-defined.

Let $x$ be a finite configuration,
$$ \sum_{j \in \mathbb{Z}} H_{\theta}(x)_j = \sum_{j \in \mathbb{Z}} \theta_j(x_{[j-r,j+r]}) $$
$$ = \sum_{j \in \mathbb{Z}} \left( x_j + \sum_{i=0}^{2r-1} \theta_{j+i+1-2r}(0^{2r-i}x_{[j-r+1,j-r+i+1]}) - \theta_{j+i-2r}(0^{2r-i}x_{[j-r,j-r+i]}) \right) $$
$$ = \sum_{j \in \mathbb{Z}} x_j + \sum_{i=0}^{2r-1} \left( \sum_{j \in \mathbb{Z}} \theta_{j+i+1-2r}(0^{2r-i}x_{[j-r+1,j-r+i+1]}) - \sum_{j \in \mathbb{Z}} \theta_{j+i-2r}(0^{2r-i}x_{[j-r,j-r+i]}) \right) $$
but
$$\sum_{j \in \mathbb{Z}} \theta_{j+i+1-2r}(0^{2r-i}x_{[j-r+1,j-r+i+1]}) = \sum_{j \in \mathbb{Z}} \theta_{j+i-2r}(0^{2r-i}x_{[j-r,j-r+i]})$$
then
$$\mu(H_{\theta}(x)) = \sum_{j \in \mathbb{Z}} H_{\theta}(x)_j = \sum_{j \in \mathbb{Z}} x_j = \mu(x)$$
and $H_{\theta}$ is FNC then NC because it is a r$\nu$-CA.
\end{proof}

\begin{theorem}
$NC(\mathcal{R})$ is a subshift of finite type.
\end{theorem}

\begin{proof}
$NC(\mathcal{R})$ is the set of distributions which avoid the pattern of $\mathcal{F}_{\mathcal{R}}$ which is finite as a subset of $\mathcal{R}^{2r+1}$. 
\end{proof}

\section{Equicontinuity and sensitivity for linear r$\nu$-CA}

\label{eq_sens_add}

In this part, we will have a look on dynamical properties on linear  r$\nu$-CA.  All along this section, $(A,+,.)$ denotes a finite commutative ring, $0$ and $1$ denote the neutral elements of $(A,+)$ and $(A,.)$, respectively.

Then for all integer $n$, $(A^n, +, .)$ defines an A-algebra by
\begin{enumerate}
\item
$\forall u,v \in A^n, u + v = (u_0 + v_0, \dots, u_{n-1} + v_{n-1})$
\item
$\forall u,v \in A^n, uv = (u_0v_0, \dots, u_{n-1}v_{n-1})$
\item
$\forall \lambda \in A, \forall u \in A^n, \lambda u = (\lambda u_0, \dots, \lambda u_{n-1})$
\end{enumerate}

Similarly $(\mathcal{C}_A, +, .)$ defines an A-algebra by
\begin{enumerate}
\item
$\forall x,y \in \mathcal{C}_A, \forall i \in \mathbb{Z}, (x + y)_i = x_i + y_i$
\item
$\forall x,y \in \mathcal{C}_A, \forall i \in \mathbb{Z}, (x y)_i = x_i  y_i$
\item
$\forall \lambda \in A, \forall x \in \mathcal{C}_A, \forall i \in \mathbb{Z}, (\lambda x)_i = \lambda x_i$
\end{enumerate}

A $\nu$-CA $H : \mathcal{C}_A \rightarrow \mathcal{C}_A$ is said to be \emph{linear} if and only if for all configurations $x$ and $y$, $H(x+y) = H(x) + H(y)$. Similarly a local rule $f$ of radius $r$ is said to be linear if and only if for all words $u$ and $v$ in $A^{2r+1}$, $f(u+v)  = f(u) + f(v)$.

A local rule $f$ of radius $r$ is linear if and only if there exists a word $\lambda$ in $A^{2r+1}$ such that
$$\forall u \in A^{2r+1}, f(u) = \lambda \bullet u := \sum_{i=0}^{2r} \lambda_i u_i$$

A $\nu$-CA $H : \mathcal{C}_A \rightarrow \mathcal{C}_A$ is linear if and only if
$$\forall i \in \mathbb{Z}, \exists r \in \mathbb{N}, \exists f \in \mathcal{R}_{A,r}, \forall x \in \mathcal{C}_A, H(x)_i = f(x_{[i-r,i+r]})$$
and $f$ is linear for all integer $i$.

We will  be interested in equicontinuity and sensitivity for linear  $\nu$-CA. In the general case, a $\nu$-CA is not sensitive if and only if it admits an equicontinuous point. In the case of additive $\nu$-CA a stronger property holds.

\begin{proposition}
Let $H$ be a linear $\nu$-CA then $H$ is either sensitive or equicontinuous.
\end{proposition}

\begin{proof}
$H$ is linear, then for all integer $n$, $H^n$ is linear. For all integers $n$ and $i$, there exists an integer $r \geq 0$  and $\lambda \in A^{2r+1}$ such that for all configuration $x$, $H^{n}(x)_i = \lambda \bullet x_{[i-r,i+r]}$. Let $r_i^n$ denotes $\max (\{i > 0 : \lambda_{r - i} \neq 0 \text{ or }  \lambda_{r + i} \neq 0\} \cup \{0\})$. It is easy to see that $r_i^n$ is well-defined whatever the choice of $r$ and $\lambda$ is done.

Assume there exists $i$ such that the sequence $(r_i^n)_{n \in \mathbb{N}}$ is not bounded. Let $x$ be a configuration, let $\delta = 2^{-m} > 0$, there exists an integer $n$ such that $r_i^n > 2|i| + 1 + m$, let $y$ be the configuration such that for all integer $j$, $y_j = 0$ if $|j| \neq r_i^n$ ; 1 otherwise. Then $d(x,x+y) = d(0,y) < \delta$ and $d(H^n(x), H^n(x+y)) = d(0, H^n(y)) > 2^{-i}$. Then $H$ is sensitive with sensitivity constant $2^{-i}$.

Assume at opposite that for all integer $i$ the sequence $(r_i^n)_{n \in \mathbb{N}}$ is bounded by the integer $M_i > 0$. Let $x$ be a configuration, let $\epsilon = 2^{-m} > 0$, let $\delta = 2^{-(m + M)}$ where $M = \max \{M_i : -m \leq i \leq m\}$, let $y$ be a configuration such that $d(x,y) < \delta$, then for all integer $n$, $d(H^n(x) = H^n(y)) < \epsilon$ because $x_{[-m-M, m + M]} = y_{[-m-M, m + M]} \rightarrow H^n(x)_{[-m,m]} = H^n(y)_{[-m,m]}$. Then $H$ is equicontinuous.

Then $H$ is either sensitive or equicontinuous.
\end{proof}

From now on $\mathcal{R}$ is a finite set  of linear local rules of radius $r$.

\begin{definition}[Wall]
Let $\psi \in \mathcal{R}^*$ of size $n \geq r$, then $\psi$ is a right-wall if and only if all the sequences $(u_k)_{k \in \mathbb{N}}$ defined by
$$
\begin{array}{rcl}
u_0 & = & 0^n \\
u_1 & = & h_{\psi}(0^ru_0v) \text{ where $v \in A^r$} \\
u_{k+1} & = & h_{\psi}(0^ru_k0^r) \text{ for $k > 1$}
\end{array}
$$
verify ${u_k}_{[0,r-1]} = 0^r$. Left-wall are defined similarly. 
\end{definition}

\begin{proposition}
Let $\theta \in \Theta_{\mathcal{R}}$. Then, $H_{\theta}$ is sensitive if and only if there exist two integers $k^-$ and $k^+$ such that
\begin{enumerate}
\item
for all integer $i < k^-$, for all integer $n \geq r - 1$, $\theta_{[i - n,i]}$ is not a left-wall
\item
for all integer $i > k^+$, for all integer $n \geq r - 1$, $\theta_{[i,i+n]}$ is not a right-wall
\end{enumerate}
\end{proposition}

\begin{proposition}
If all the rules of $\mathcal{R}$ have radius 1 then the language $\{ \theta \in \Theta_{\mathcal{R}} : H_\theta \text{ is sensitive}\}$ is a $\zeta$-rational language.
\end{proposition}

\bibliographystyle{acm}
\bibliography{rnuca}

\end{document}